\documentclass[lettersize,journal]{IEEEtran}
\usepackage{lineno,hyperref}
\usepackage{float}
\usepackage{graphicx}
\usepackage{epstopdf}
\usepackage{algorithm}
\usepackage{algorithmicx}
\usepackage{algpseudocode}
\usepackage{amsmath}
\usepackage{amstext}
\usepackage{amssymb}
\usepackage{caption}
\usepackage{subfigure}
\usepackage{color}
\usepackage{multirow}
\usepackage{subfigure}
\usepackage{listings}
\usepackage{xcolor}
\usepackage[numbers,sort&compress]{natbib}
\usepackage{setspace}
\usepackage{makecell}
\newcommand{\lin}{\textcolor{black}}

\begin{document}

\title{A Unified Framework for Integrating Semantic Communication and AI-Generated Content in Metaverse}

\author{
        Yijing Lin,
        Zhipeng Gao,
        Hongyang Du,
	Dusit Niyato,~\IEEEmembership{Fellow,~IEEE} \\
        Jiawen Kang,
        Abbas~Jamalipour,~\IEEEmembership{Fellow,~IEEE},
        and Xuemin Sherman Shen,~\IEEEmembership{Fellow,~IEEE}
\thanks{This work is supported by National Natural Science Foundation of China (62072049). Corresponding author: Zhipeng Gao (e-mail: gaozhipeng@bupt.edu.cn). Yijing Lin and Zhipeng Gao are with the State Key Laboratory of Networking and Switching Technology, Beijing University of Posts and Telecommunications, China (e-mail: yjlin@bupt.edu.cn; gaozhipeng@bupt.edu.cn). Hongyang Du and Dusit Niyato are with the School of Computer Science and Engineering, Nanyang Technological University, Singapore (e-mail: hongyang001@e.ntu.edu.sg; dniyato@ntu.edu.sg). Jiawen Kang is with Guangdong University of Technology, 510006, Guangzhou, China (e-mail: kjwx886@163.com). Abbas Jamalipour is with The University of Sydney, Sydney NSW 2006, Australia (e-mail: a.jamalipour@ieee.org). Xuemin Sherman Shen is with the Department of Electrical and Computer Engineering, University of Waterloo, Canada (e-mail: sshen@uwaterloo.ca).}
}

\maketitle

\begin{abstract}

As the Metaverse continues to grow, the need for efficient communication and intelligent content generation becomes increasingly important. Semantic communication focuses on conveying meaning and understanding from user inputs, while AI-Generated Content utilizes artificial intelligence to create digital content and experiences. Integrated Semantic Communication and AI-Generated Content (ISGC) has attracted a lot of attentions recently, which transfers semantic information from user inputs, generates digital content, and renders graphics for Metaverse. In this paper, we introduce a unified framework that captures ISGC’s two primary benefits: integration gain for optimized resource allocation and coordination gain for goal-oriented high-quality content generation to improve immersion from both communication and content perspectives. We also classify existing ISGC solutions, analyze the major components of ISGC, and present several use cases. We then construct a case study based on the diffusion model to identify a \lin{near-optimal} resource allocation strategy for performing semantic extraction, content generation, and graphic rendering in the Metaverse. Finally, we discuss several open research issues, encouraging further exploring the potential of ISGC and its related applications in the Metaverse.
\end{abstract}


\section{Introduction}
\IEEEPARstart{T}{he} concept of Metaverse, originally introduced in the scientific novel Snow Crash, has attracted considerable interest from academia and industries. Metaverse refers to a virtual environment that seamlessly integrates with the physical world, allowing for the existence of digital avatars \lin{to engage in various activities, interact with other users, and access virtual objects and experiences. The construction of Metaverse is supported by all virtual reality (VR), augmented reality (AR), and the Internet of Things to create a comprehensive and interconnected digital ecosystem.}

The continuous advancement of technologies such as semantic communication (SemCom) and AI-Generated Content (AIGC) has prompted the Metaverse to increase demands for efficient communication and intelligent content generation. \lin{Semantic communication refers to focusing on the associated meanings rather than simply transmitting raw data to enable effective communication. AIGC generates digital content automatically with the assistance of AI technologies to improve efficiency and provide personalized and relevant content tailored to the preferences and needs of the users. These technologies lead to} the emergence of a new integration technology: integrated SemCom and AIGC (ISGC) for improving immersion from both communication and content perspectives. \lin{ISGC combines the benefits of SemCom and AIGC to extract autonomously relevant information from raw data, enabling the generation of high-quality digital content in the Metaverse without direct human intervention. Additionally,} There are certain difficulties that may arise without tight integration between SemCom and AIGC for Metaverses:

\begin{itemize}
    \item \textbf{Inefficient Use of Resources:} \lin{Recognizing the challenges of the collaborative execution of AIGC tasks across a multitude of devices and the diverse access requirements of users \cite{du2023exploring},} there is a lack of integration in the allocation of computing and communication resources for semantic extraction, AIGC, and graphic rendering tasks, leading to \lin{near-optimal} resource utilization and \lin{performance}.
    \item \textbf{Low-quality of Content:} Without effective coordination between SemCom and AIGC, the generated content may not meet the desired quality standards \lin{\cite{huang2021deep}}. This can lead to poor user experiences and dissatisfaction, ultimately affecting the adoption and success of Metaverse.
\end{itemize}

As a consequence, ISGC has emerged as a promising technology that combines the advantages of the aforementioned technologies. \lin{By combining AIGC and SemCom, ISGC enables the production of content that is not only visually appealing but also contextually relevant and meaningful, enhancing user experiences in the Metaverse. It also ensures that the right resources are allocated to the right tasks at the right time through joint computing and communication resource optimization. Additionally, it can adapt to user preferences, contextual information, and real-time interactions.} \lin{In summary}, ISGC can provide two main benefits over the above functionalities by obtaining the integration and coordination gains \cite{cui2021integrating} \lin{for optimized resource allocation and high-quality content generation}. 

This paper provides a conceptual overview and concrete use cases of ISGC as well as its role in the Metaverse. Specifically, we present the related works and the key benefits of ISGC. We propose a unified framework for ISGC that utilizes the advantages of integration and coordination gains. Furthermore, we provide a case study employing the diffusion model to determine the \lin{near-optimal} strategy for resource allocation in ISGC. It is shown that the diffusion model is capable of handling the effects of randomness and noise, and promoting exploration to enhance policy flexibility. To the best of our knowledge, this work is the first to comprehensively explore the potential integration and coordination benefits of ISGC for improving the efficiency and intelligence of the Metaverse. Our main contributions are summarized as follows:

\begin{itemize}
    \item We propose a comprehensive overview of ISGC, including an investigation of related works, an explanation of why SemCom and AIGC integration is necessary, and the reasons why ISGC is needed in the Metaverse.
    \item We present a unified framework for ISGC, which includes a step-by-step workflow for capturing the integration and coordination gains, as well as several potential use cases.
    \item To further explore the benefits of integration gains, we conduct a case study that analyzes the effects of ISGC on resource allocation. Specifically, we use the diffusion model to derive and promote \lin{near-optimal} strategies for utilities of resource allocation.
\end{itemize}

\section{Why Integrate SemCom and AIGC}
\label{sec_framework}

\begin{figure*}[!t]
    \centering
    \includegraphics[width=1\textwidth]{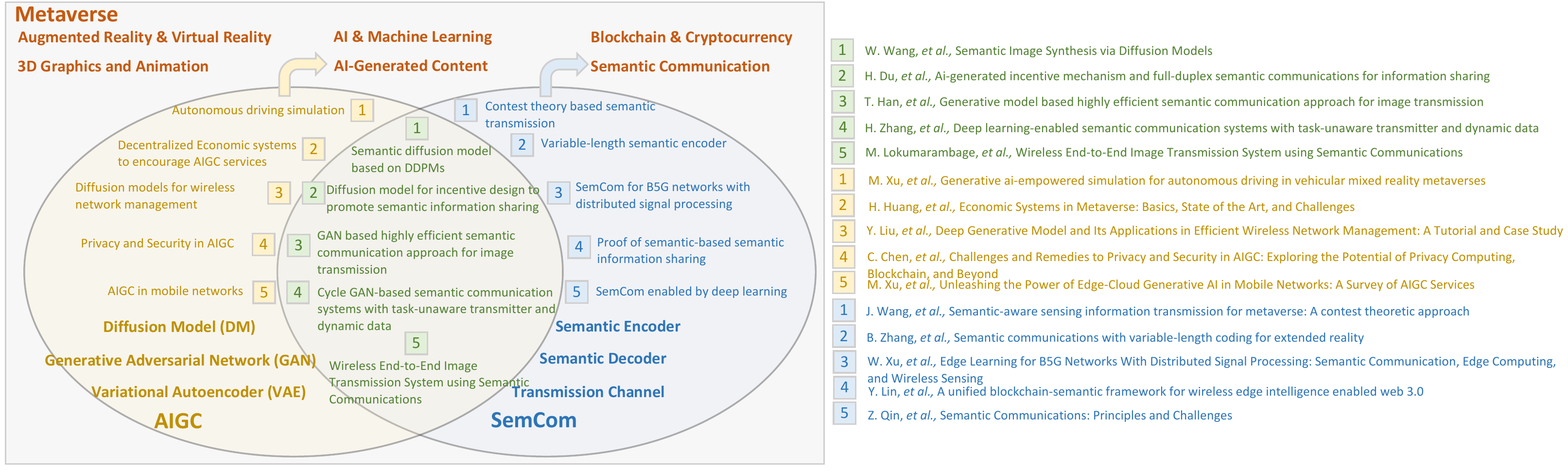}
    \caption{\lin{A review of recent research studies and emerging trends across SemCom, AIGC, and Metaverse, inspiring a unified framework for integrated SemCom and AIGC in the Metaverse}}
    \label{fig_reference}
\end{figure*}

\lin{\textit{\textbf{ISGC}} is a design paradigm in which SemCom and AIGC are integrated to provide efficient communication and goal-oriented content generation.} A notable surge in research activities pertaining to ISGC has been observed, as shown in Fig. \ref{fig_reference}. We collect several papers \textit{from IEEE Xplore and arXiv in April 2023} \lin{and identify research trends and directions from Fig.\ref{fig_reference} that depict the current research of ISGC.}

\begin{figure}[!t]
    \centering
    \includegraphics[width=0.5\textwidth]{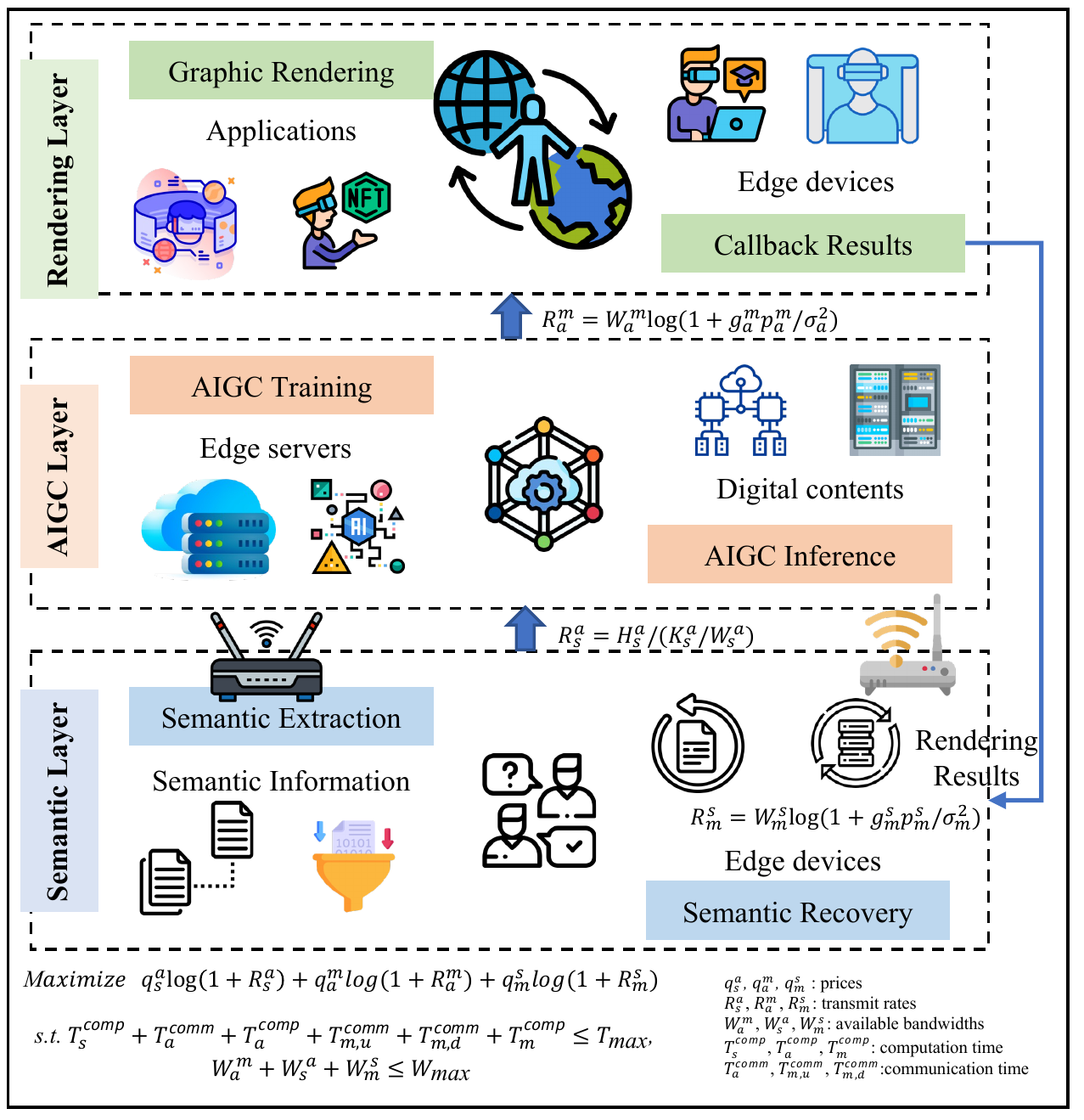}
    \caption{\lin{A layered view of ISGC-aided Metaverse: Semantic, AIGC, and Rendering layers}}
    \label{fig_equation}
\end{figure}

\textit{1) SemCom and AIGC.} The integration of SemCom and AIGC primarily aims to leverage AIGC technologies such as Generative Adversarial Networks to develop semantic decoders that address the out-of-distribution problem between transmitters and receivers \cite{zhang2022deep}. To compute the loss function, a variational autoencoder is employed to calculate the lower bound of semantic distortion, while diffusion models are combined with deep reinforcement learning to determine the \lin{near-optimal} decisions in semantic communication \cite{alemi2016deep}. 

\textit{2) SemCom and Metaverse.} \lin{The integration of SemCom and Metaverse aims to circulate meaningful information with fewer symbols within the Metaverse, thereby reducing communication overheads. In order to mitigate privacy concerns arising from this integration, federated learning is introduced to preserve user data privacy \cite{chen2023trust}}.

\textit{3) AIGC and Metaverse.} To improve the integration of AIGC and Metaverse, the focus is on generating high-quality digital content to create immersive virtual environments and construct economic systems, such as autonomous driving simulations and customized content. Additionally, the integration utilizes diffusion models to efficiently manage and optimize network and resource allocation \cite{xu2023generative}. 

\textit{4) SemCom, AIGC and Metaverse.} The integration of SemCom, AIGC, and Metaverse is still in its early stages and is primarily focused on improving the efficiency of Metaverse through the application of SemCom and AIGC. GAN is utilized for the extraction of semantic information to improve the transmission efficiency in Metaverse \cite{han2022generative}.

\textit{\textbf{Layers of Integration.}} Fig. \ref{fig_equation} depicts the various layers involved in the integration of ISGC. \lin{Data collected from sensors} is extracted and transformed into semantic information such as image segments or model features, which is transmitted using semantic communication. AIGC inference is then applied to generate digital content from this information. The generated content is then fused through rendering graphics to create virtual environments that can be used by various applications and users within \lin{the Metaverse ecosystem.}

\textit{\textbf{Research Trends.}} To identify research trends related to ISGC, the research activities are classified into different layers based on their integration, as shown in \lin{Table} \ref{fig_trend}. The figure highlights that current research solutions predominantly focus on addressing issues caused by individual layers, such as \lin{the out-of-distribution problem between the semantic encoder and decoder, efficient incentive mechanism for sharing semantic information, decentralized semantic sharing, and resource allocation for tasks within the same layers.} However, these solutions may not fully reflect the benefits of the integration, and there is a need for more research efforts to explore the potential of ISGC as a whole.

\begin{table*}[!t]
    \centering
    \includegraphics[width=1\textwidth]{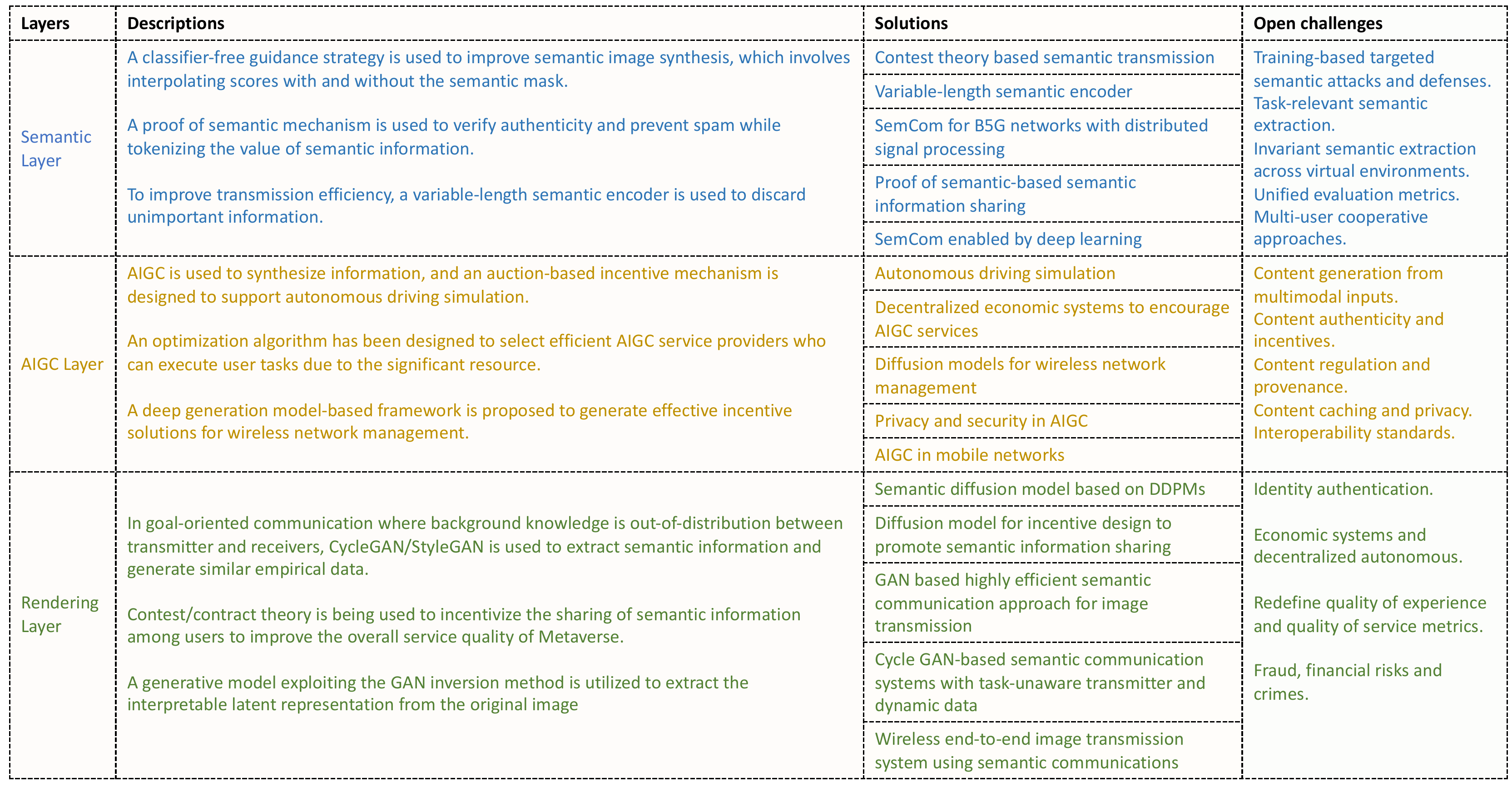}
    \caption{\lin{ISGC-aided Metaverse: solutions, descriptions, and their integration challenges}}
    \label{fig_trend}
\end{table*}

\section{ISGC Unified Framework}

\subsection{Framework Overview}

ISGC includes semantic, inference, and rendering modules to capture the benefits of the integration of SemCom, AIGC, and Metaverse.

\textit{1) Semantic Module:} To optimize the data processing stage and reduce communication overhead, data collection, data processing, and semantic extraction should be performed at the edge devices. The semantic module is specifically designed to process data generated by edge devices and extract semantic information from raw data simultaneously. The extracted semantic information is then transmitted to MSPs that control the AIGC and render modules via edge servers. 

\textit{2) Inference Module:} Semantic information is fed into semantic decoder to recover useful information. Since the recovered images are low-quality or incomplete, MSPs should utilize AIGC to generate high-quality digital content to improve user experiences. The inference module employs pre-trained models to generate high-quality images with depth maps from multiple angles via the latent diffusion model, which employs forward and reverse diffusion processes to add and remove noise from images.

\textit{3) Rendering Module:} Empowered by the above modules, the render module can synthesize massive and conditioned information from real-world or imaginary scenarios to enable immersive and interactive virtual environments.

\subsection{Major Issues in Separated Functionalities}

When the functionalities of SemCom, AIGC, and Metaverse are separated without ISGC, several significant issues may arise.

\textit{1) Resource Underutilization:} Current resource allocation solutions tend to focus on individual modules, rather than considering the integrated ISGC as a whole. For instance, J. Wang \textit{et al.} \cite{wang2023semantic} proposed using contest theory to incentive users in the semantic module to contribute more valuable information. However, this approach may lead to the overuse of certain resources in one module while leaving others idle, resulting in inefficient resource allocation and decreased performance.

\textit{2) Limited Flexibility:} The information provided by individual modules may suffer from high transmission latency or be affected by noise from the communication channel, leading to a decrease in the quality of user experience in Metaverse. For instance, B. Zhang \textit{et al.} \cite{zhang2023semantic} proposed a variable-length channel coding method to highly compress \lin{unimportant semantic information} to improve transmission efficiency. However, this approach may generate low-quality content in Metaverse. Additionally, the content generated by AIGC within Metaverse may require meaningful information from users to enhance the quality of particular applications.

\subsection{\lin{Technical Gains} of ISGC in Metaverse}

\textit{\textbf{Integration Gain.}} \lin{It can be achieved through resource allocation and sharing, specifically in terms of computing, communication, and dataset sharing among SemCom, AIGC, and Metaverse. A strategic allocation or balance of resources can be implemented based on environmental conditions and user requirements to maximize overall utilities. In cases where the channel conditions are unfavorable, it becomes impractical to allocate excessive resources to AIGC and the Metaverse, as they would be limited by the performance of SemCom. Instead, allocating more resources to SemCom can yield optimal utilities. This process can be seen as maximizing minimum utilities among SemCom, AIGC, and Metaverse.} The workflow for dynamically coupling resources of ISGC consists of the following three steps. More details are shown in the next section.

\begin{itemize}
    \item \textbf{Step 1: Design the joint optimization problem.} Given computing and communication resources, ISGC needs to consider both the usage of resources and latency in each module. To achieve this, ISGC can construct joint resource allocation optimization problems to maximize utility.
    \item \textbf{Step 2: Learn the policy via training.} Diffusion model-based Deep Reinforcement Learning (DRL) is utilized to solve the joint optimization problem and learn the policy since the diffusion model could mitigate the effect of randomness and noises \cite{du2023ai}.
    \item \textbf{Step 3: Generate the \lin{near-optimal} strategy via inference.} The trained model can generate \lin{near-optimal} strategies based on dynamic inputs to improve efficiency of the integration.
\end{itemize}

\begin{figure*}[!t]
    \centering
    \includegraphics[width=1\textwidth]{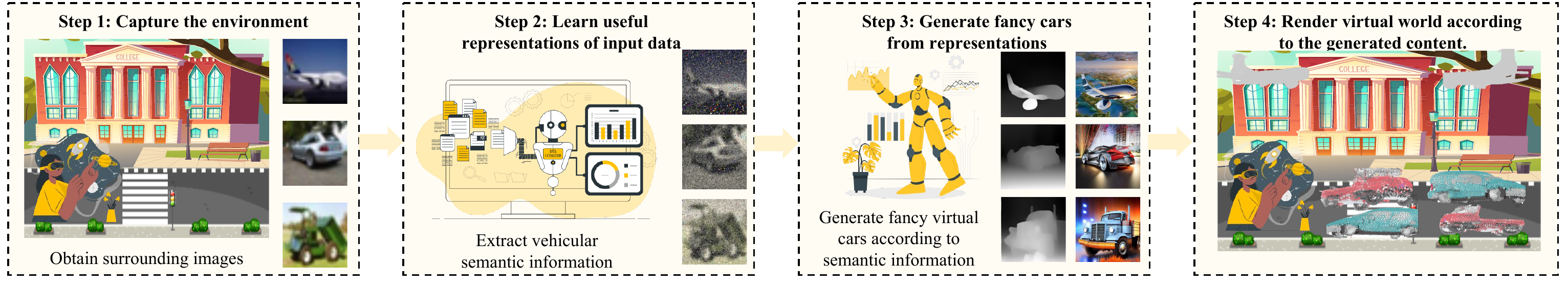}
    \caption{An illustration of the coordination gain of ISGC}
    \label{fig_case_study}
\end{figure*}

\textit{\textbf{Coordination Gain.}} The coordination gain achieved by ISGC is essential in achieving goal-oriented content generation within Metaverse, which can couple the functionalities of semantic communication, AIGC inference, and graphic rendering more tightly. For coordination gain, we can customize SemCom based on the AIGC algorithm and Metaverse user requirements. For example, if a user is participating in virtual driving, SemCom should focus on vehicular network semantic information. By integrating SemCom, AIGC, and Metaverse, ISGC can extract semantic information efficiently, generate high-quality content with AI, and seamlessly integrate it into the Metaverse ecosystem. Unlike separate functionalities, the integration of ISGC ensures to perform mutual assistance. To provide a more concrete illustration of the coordination gain achieved by ISGC, a use case involving a virtual campus is presented in Fig. \ref{fig_case_study}.

\begin{itemize}
    \item \textbf{Step 1: Capture the environment.} In the scenario of a university campus, sensors, e.g., camera sectors of VR/AR/XR devices, capture the environmental settings from the real campus, like animals running around or airplane flying \lin{in} the sky.
    \item \textbf{Step 2: Learn useful representations of input data.} Semantic information, e.g., feature vectors, is extracted from images in the semantic modules and transmitted to the inference module controlled by MSPs.
    \item \textbf{Step 3: Generate depth maps from representations.} MSPs first use feature vectors to reconstruct low-quality images and then generate depth maps with multiple angles of the environmental settings in the inference module.
    \item \textbf{Step 4: Render virtual campus with personalized feedback from depth maps.} The rendering model could provide personalized feedback to devices based on the above depth maps to simulate real-world settings.
\end{itemize}

\section{Case Study: Exploring Integration Gain}

Within the ISGC framework, to explore the integration gain, given limited available computing and communication resources, they need to be allocated to semantic extraction, AIGC inference, and graphic rendering modules to maximize the end-to-end utility. In this section, we first formulate the utility joint optimization problem of ISGC, as shown in Fig. \ref{fig_equation}, then achieve an effective resource allocation mechanism to obtain \lin{near-optimal} strategies, and depict the simulation results of the proposed mechanism.

\subsection{Problem Formulation}

\lin{To simplify the notation, we use subscripts and superscripts of $s, a, \mbox{ and } m$ to represent the semantic, AIGC, and rendering modules, respectively. Additionally, $\mathsf{comm}$ represents the communication time, and $\mathsf{comp}$ represents the computation time.}

\textit{1) Semantic Extraction.} Edge devices employ the semantic module to extract semantic information \lin{from raw data}, which reduces the amount of data transmitted by using fewer symbols. As mentioned in \cite{liu2019deep}, the computation time $T_s^{\mathsf{comp}}$ for semantic extraction depends on the available computational resources of edge devices. Specifically, it is determined by the ratio of the required computational resources $Z_s$ of semantic extraction to the total resources available $C_s$ on edge devices. 

\textit{2) Semantic Module to Inference Module.} The semantic rate $R_s^a$ refers to the amount of semantic information transmitted per second \cite{yan2022qoe}. It is determined by considering the proportion of the approximate semantic entropy $H_s^a$, \lin{a measure of the uncertainty or randomness associated with semantic information}, to the available bandwidth $W_s^a$ between edge devices and MSPs and the average number of transmitted symbols $K_s^a$. Consequently, the time $T_a^{\mathsf{comm}}$ required to transmit semantic information $D_s$ from the semantic module on edge devices to the inference module on edge servers is the ratio of the extracted semantic information to the semantic rate $R_s^a$. 

\textit{3) AIGC Inference.} Upon receiving semantic information from edge devices, MSPs carry out AIGC inference tasks that are guided by the semantic information to conditionally generate digital content in edge servers. The time required for AIGC inference $T_a^{\mathsf{comp}}$ is influenced by the computational resources available on edge servers containing the inference module. In particular, this time is dictated by the proportion of the necessary computational resources $Z_a$ for AIGC inference to the overall resources $C_a^s$ present on edge servers managed by MSPs. 

\textit{4) Inference module to Rendering module.} The transmission rate $R_a^m$ from the inference module to the rendering module is computed as the product of the bandwidth available $W_a^m$ between edge servers and MSPs and the channel capacity, referring to \cite{xu2023generative}. The channel capacity is influenced by the channel gain $g_a^m$, transmit power $p_a^m$, and the additive Gaussian noise $\sigma_a^2$. The transmission time $T_{m,u}^{\mathsf{comm}}$ is determined by the ratio of the data size of the generated AIGC digital content to the transmission rate. 

\textit{5) Graphics Rendering.} Once the digital content is received from the corresponding edge servers running the inference module, MSPs equipped with the rendering module undertake graphics rendering tasks. These tasks involve leveraging digital content to augment and enrich virtual environments. The computation time $T_m^{\mathsf{comp}}$ required for graphics rendering is contingent upon the available computational resources of the edge servers deploying the rendering module. Precisely, this time is derived from the proportion of the required computing resources $Z_m$ to the aggregate resources accessible $C_m^a$ on these edge servers.

\textit{6) Rendering module to Users.} The transmission rate $R_m^s$ between the rendering module and the semantic module can be analogous to that between the AIGC and rendering module \lin{with the bandwidth available $W_m^s$, channel gain $g_m^s$, transmit power $p_m^s$, and the additive Gaussian noise $\sigma_m^2$}. The transmission time $T_{m,d}^{\mathsf{comm}}$ is determined by the ratio of the data size of the rendering feedback to the transmission rate.

\textit{7) MSP Utility.} MSPs impose charges on edge devices for the transmission and execution of tasks on edge servers. Referring to \cite{liu2019deep}\cite{yan2022qoe}, the utility of MSPs can be determined by the products of the price $q_s^a, q_a^m, q_m^s$ and transmit rate $R_s^a, R_a^m, R_m^s$ of \lin{semantic, AIGC, and rendering modules}. The utility is limited by the tolerable transmission time and the given bandwidth resources among the three modules, as shown in Fig. \ref{fig_equation}. 

\subsection{\lin{Diffusion Model-Based Joint Resource Allocation}}

Inspired by \cite{wang2022diffusion}, in this paper, we present the diffusion model-based joint resource allocation mechanism. This mechanism is characterized as a Markov decision process consisting of state spaces, action spaces, environment dynamics, a reward function, a discount factor, and an initial state distribution. \lin{The reward is calculated by the utility function.} The primary objective of this mechanism is to learn a policy that maximizes the cumulative discounted reward, thereby optimizing the utilities for MSPs within the ISGC framework. 

\textit{\textbf{AI-Generated Resource Allocation:}} \lin{The resource (i.e., bandwidth) allocation problem is solved by the diffusion model, which is composed of forward and reverse processes. The processes are designed to add and remove noises from samples, ultimately yielding generative outputs.} The diffusion model can be further extended to include conditional models to represent the policy that optimizes the rewards for MSPs \cite{wang2022diffusion}. \lin{The conditional diffusion model is integrated with DRL to iteratively denoise the initial distribution and produce a near-optimal utility function for MSPs.}

\begin{itemize}
    \item \textbf{Step 1: Design state spaces.} Based on the MSPs' utility derived in the previous section, the \lin{near-optimal} strategy $\pi(a^0|s \in \mathcal{S})$ is influenced by a variety of factors, denoted as \textit{state spaces} $\mathcal{S}$. These state spaces $[H_s^a, \sigma_a, \sigma_m, g_a^m, p_a^m, g_m^s, p_m^s,  K_s^a, C_a^s, C_m^a]$ include the approximate semantic entropy and the average transmitted symbols of the semantic module, channel gains and transmit power from the inference module to the rendering module, channel gains and transmit power from the rendering module to the inference module, as well as the computing resources and additive Gaussian noises present at the AIGC and rendering modules.
    \item \textbf{Step 2: Construct action spaces.}  Given the state spaces, the \textit{action spaces} $a^0 \in \mathcal{A}$ are associated with several factors, including the available bandwidth from the semantic, AIGC, and rendering modules, respectively. Consequently, the diffusion model that establishes a mapping between states $\mathcal{S}$ as the condition and action $\mathcal{A}$ as outputs represent the \lin{near-optimal} policy $\pi(a^0|s \in \mathcal{S})$. This policy yields a deterministic resource allocation strategy, which aims to maximize the expected cumulative reward over a series of steps.

    \item \textbf{\lin{Step 3: Explore the training policy in the forward process.}} Initiating the training step involves providing hyper-parameters, including diffusion steps $T$, batch size, and exploration noise. The diffusion model is then initialized, incorporating two critic networks along with their corresponding target networks with different weights. In each iteration, the method initializes a random Gaussian distribution $c^T$ for resource allocation exploration, followed by entering a loop of multiple steps. During each step, the method initially observes the current environment and its associated states, then sets the current actions as Gaussian noise. Subsequently, it generates the next action by denoising the current action $p(a^{i}|a^{i+1},s)$ through the reverse diffusion process and adds exploration noise to the generated action. Once the action is executed, the method obtains the corresponding reward based on the utility function and stores the environment record in the replay buffer. To further refine the model, it samples a random mini-batch of records from the replay buffer, updates the critic networks by computing the loss and policy gradient, and finally updates the target networks.

    \item \textbf{\lin{Step 4: Generate near-optimal resource allocation strategy in the reverse process.}} In the inference step, the environment with its associated states is input into the networks. Subsequently, the \lin{near-optimal} resource allocation strategy $\pi(a^0|s \in \mathcal{S})$ is generated by denoising Gaussian noise through the reverse diffusion process. This step focuses on utilizing the trained model to produce effective resource allocation strategies based on the given environmental conditions.
\end{itemize}

\subsection{Simulation Results}

The experimental platform utilized for executing the \lin{bandwidth} resource allocation is built on a generic Ubuntu 20.04 system, featuring an AMD Ryzen Threadripper PRO 3975WX 32-Core CPU and an NVIDIA RTX A5000 GPU. The approximate semantic entropy, average transmitted symbols, channel gain and transmit power between the AIGC and rendering modules, as well as channel gain and transmit power between edge servers and devices, are randomly sampled from uniform distributions (1, 2), (0, 0.8), (0, 1), (3, 5), (0, 1), and (3, 5), respectively. The additive Gaussian noise at the AIGC and rendering modules is randomly sampled from normal distributions (0, 1) and (0, 1), respectively. The constraints of the total interaction time, the available bandwidth between the semantic, AIGC, and rendering modules, respectively. The above parameters are set as indicated in \cite{yan2022qoe,liu2019deep,xu2023generative}. 

\lin{In the simulation experiment, the diffusion model (Diffusion) and the Proximal Policy Optimization (PPO) \cite{schulman2017proximal} algorithm with learning rates, 3e-7 and 3e-6 are used to determine the near-optimal allocation of the available bandwidth among the semantic module, AIGC module, and rendering module.} Unless otherwise specified, these methods are assumed to operate under identical parameters and environments. PPO is a model-free, on-policy actor-critic algorithm that uses the clipped surrogate objective to improve the stability and efficiency of learning. The training process is set to run for 3,000 epochs with buffer size \lin{1,000,000}, exploration noise 0.01, 10 steps per epoch, and 100 steps per collect, providing sufficient iterations for these methods to learn and adapt to the given environment and parameters.

\begin{figure}[!t]
    \centering
    \includegraphics[width=0.45\textwidth]{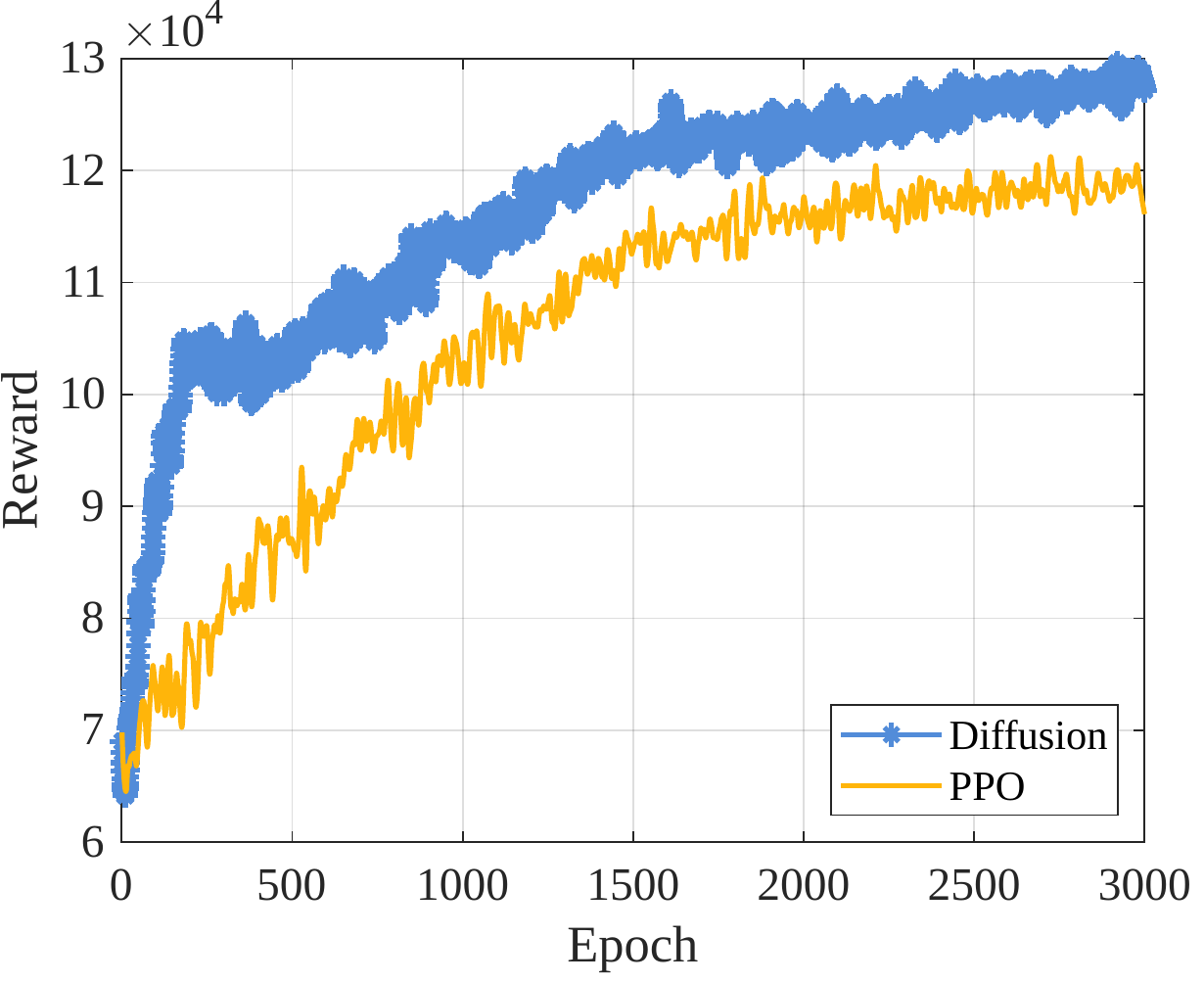}
    \caption{\lin{Training curves of the joint resource allocation}}
    \label{fig_diffusion_reward}
\end{figure}

\begin{figure}[!t]
    \centering
    \includegraphics[width=0.45\textwidth]{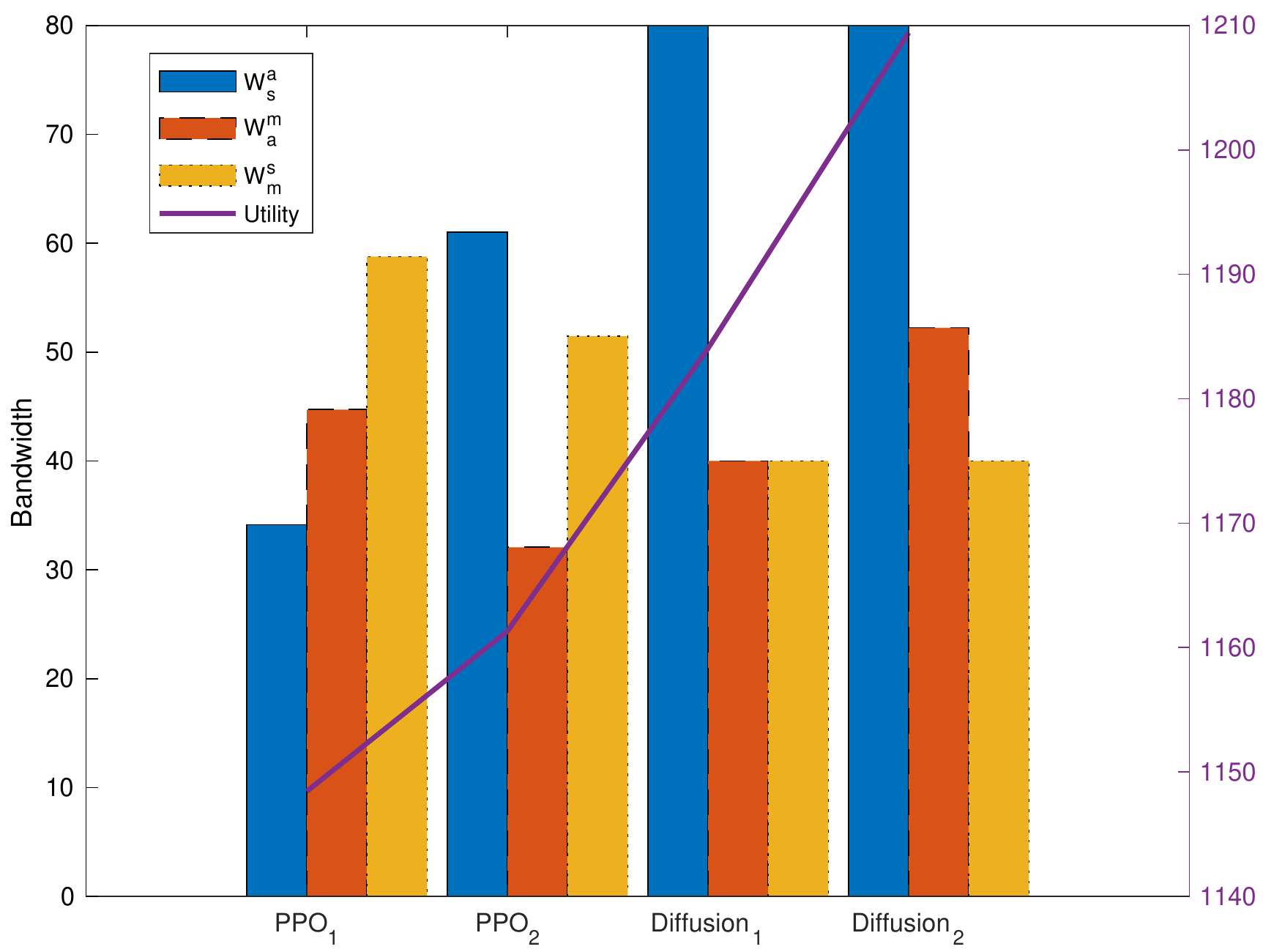}
    \caption{\lin{Generated utility of Diffusion compared with PPO}}
    \label{fig_diffusion_utility}
\end{figure}

Fig.~\ref{fig_diffusion_reward} illustrates the reward comparison among our proposed mechanism and PPO. The training process demonstrates the reward values of Diffusion are significantly higher than those of PPO. Fig.~\ref{fig_diffusion_utility} compares the utilities computed by the \lin{near-optimal} actions \lin{under different network states $[H_s^a, \sigma_a, \sigma_m, g_a^m, p_a^m, g_m^s, p_m^s,  K_s^a, C_a^s, C_m^a]$, i.e., $\mathsf{PPO}_1$ with [1.17, 0.66, 1.97, 0.30, 0.24, 4.76, 4.46, 0.91, 8.03, 15.28], $\mathsf{PPO}_2$ with [1.40, 0.30, 0.65, 0.58, 0.16, 3.42, 4.69, 0.89, 7.49, 15.24], $\mathsf{Diffusion}_1$ with [1.80, 1.05, 0.47, 0.05, 0.0004, 4.23, 4.58, 0.91, 5.02, 19.73], and $\mathsf{Diffusion}_2$ with [1.52, 0.12, 1.23, 0.03, 0.14, 4.83, 3.86, 0.96, 9.93, 18.42], in the Diffusion and PPO methods}. The \lin{near-optimal} strategy found by Diffusion is better than that of PPO. The underlying cause for these outcomes is that the diffusion model-based resource allocation mechanism can adapt outputs by fine-tuning given the diffusion steps and promoting exploration, thereby enhancing flexibility and mitigating the impact of uncertainty and noise encountered during the training process. This allows the proposed mechanism to achieve superior results in comparison to the other tested algorithms. 


\section{Future Directions}

Several open challenges arise from the use of ISGC in the Metaverse as shown in \lin{Table} \ref{fig_trend}. We elaborate on several of them in this section.

\paragraph{\textbf{Invariant Semantic Extraction Across Virtual Environments}} Because Metaverse may involve multiple heterogeneous devices deployed in different virtual environments, semantic extraction may unintentionally absorb irrelevant environmental information, resulting in the extraction of useless information that can cause inaccurate content generation in Metaverse. Therefore, it is crucial to consider the impacts of out-of-distribution data and extract invariant semantic information across virtual environments.

\paragraph{\textbf{Content Authenticity and Incentives}} The limited computation resources of Metaverse devices necessitate their reliance on MSPs to generate content and enable the creation of complex and resource-intensive experiences. \lin{Therefore, it is necessary to design fair incentive mechanisms to verify content authenticity and incentivize MSPs.}

\paragraph{\textbf{\lin{Practical Implementation Difficulties}}} \lin{Achieving management of computing and communication functions across different layers and service providers, poses practical challenges. Implementing even simple techniques in practice is exceedingly difficult due to limited infrastructure access, diverse deployment environments, and requirements. Overcoming these challenges necessitates addressing interoperability, resource allocation, and performance management, highlighting the complexity of integrating compute and communication functions in real-world scenarios.}

\paragraph{\textbf{\lin{Communication Security and Privacy Preserving}}} \lin{Since users should upload their semantic information for the customized AIGC-based immersive Metaverse, it is important to achieve communication security and privacy preserving. Exploring privacy-preserving AI techniques, such as federated learning, differential privacy, and secure multi-party computation, can allow for collaborative semantic extraction without exposing users' sensitive information \cite{chen2023trust}.} 

\section{Conclusion}

\lin{In conclusion, this paper has provided a comprehensive overview of ISGC in the context of the growing Metaverse. By integrating SemCom and AIGC, ISGC offers significant benefits in terms of efficient communication and intelligent content generation. The proposed unified framework captures the integration and coordination gains of ISGC, optimizing resource allocation and enhancing the quality of digital content in the Metaverse. The case study utilizing the diffusion model demonstrates an improvement of 8.3\% in rewards compared to PPO. However, there are still open research issues that need to be explored, such as privacy concerns and advanced techniques for resource allocation optimization. Overall, this paper contributes to the understanding and potential of ISGC, paving the way for immersive and intelligent experiences in the Metaverse.}

\bibliographystyle{IEEEtran} 
\bibliography{ref}
\end{document}